\documentclass[twocolumn,secnumarabic, amssymb,floatfix,
amsmath,tightenlines,showpacs,superscriptaddress,preprintnumbers,aps,prb]{revtex4-1}

\usepackage{times}
\usepackage{color}
\usepackage{graphicx}% Include figure files
\usepackage{dcolumn}% Align table columns on decimal point
\usepackage{amssymb}
\usepackage{amsmath}
\usepackage{amsfonts}
\usepackage{docs}%
\usepackage{bm}%
\usepackage[leftcaption]{sidecap}
\usepackage{wrapfig}
 \usepackage{multirow}
 \usepackage{extarrows}
 \usepackage{calc}

\usepackage[colorlinks=true,linkcolor=blue,pagecolor=blue,
filecolor=blue,menucolor=blue,urlcolor=blue,citecolor=blue,
anchorcolor=blue]{hyperref}%

\newcommand{\sns}{$S$/$N$/$S$ }
\newcommand{\sfs}{$S$/$F$/$S$ }

\newcommand{\fs}{$S$/$F$ }
\newcommand{\ns}{$S$/$N$ }
\newcommand{\n}{$N$ }

\newcommand{\s}{$S$ }

\newcommand*{\MyDef}{\mathrm{p_z=0}}
\newcommand*{\eqdefU}{\ensuremath{\mathop{\overset{\MyDef}{=}}}}% Unscaled version
\newcommand*{\eqdef}{\mathop{\overset{\MyDef}{\resizebox{\widthof{\eqdefU}}{\heightof{=}}{=}}}}

\begin{document}

\title{Long-Range Spin-Triplet Correlations and Edge Spin Currents
in Diffusive Spin-Orbit Coupled
$SNS$ Hybrids with a Single Spin-Active Interface }

\author{Mohammad Alidoust }
\email{phymalidoust@gmail.com} 
\affiliation{Department of Physics,
University of Basel, Klingelbergstrasse 82, CH-4056 Basel, Switzerland}
\affiliation{Department of Physics,
Faculty of Sciences, University of Isfahan, Hezar Jerib Avenue,
Isfahan 81746-73441, Iran}
\author{Klaus Halterman}
\email{klaus.halterman@navy.mil} \affiliation{Michelson Lab, Physics
Division, Naval Air Warfare Center, China Lake, California 93555,
USA}

\date{\today}

\begin{abstract}
Utilizing a SU(2) gauge symmetry technique in the quasiclassical diffusive
regime, we theoretically study finite-sized two-dimensional
intrinsic spin-orbit coupled
superconductor/normal-metal/superconductor ($S$/$N$/$S$) hybrid
structures
with a single spin-active interface. We consider
intrinsic spin-orbit interactions (ISOIs) that are confined
within the $N$ wire and absent in the $s$-wave superconducting electrodes
($S$). Using experimentally feasible parameters, we
demonstrate that the coupling of the ISOIs and spin moment of the
spin-active interface results in
maximum singlet-triplet conversion and accumulation of spin
current density at the corners
of the $N$ wire nearest the
spin-active interface.
By solely
modulating the superconducting phase difference,
we show
how the opposing parities
of the charge and spin currents
provide an effective venue
to experimentally examine
pure edge spin currents not accompanied by
charge currents.
These effects occur in the {\it absence} of externally imposed fields, and
moreover
are insensitive to the
arbitrary orientations of the interface spin moment.
The experimental implementation of
these robust edge phenomena are also discussed.
\end{abstract}


\maketitle

\section{Introduction}\label{sec:intro}
The interaction of a moving particles' spin with its linear momentum
embodies the so-called spin-orbit interaction (SOI). The SOI is a
quantum mechanical effect that is relativistic in origin. 
For materials possessing a strong SOI effect,
it becomes possible to manipulate 
spin currents with less dissipation, higher
speeds, and lower power consumption compared to 
conventional charge-based
devices \cite{kikkawa_1}.
Consequently, a number of 
high-performance devices
that exploit the SOI effect have been proposed, 
including, spin transistors, and devices that store or transport information \cite{wolf_1,stepanenko_1,rashba_book_1,Wunderlich_1,Miron_1}.
The types of SOIs can be categorized %in the solid-state 
as follows:
$i)$ intrinsic (originating from the electronic band structure of the
material) and $ii)$ extrinsic (originating from the spin-dependent
scattering from impurities) \cite{zhang_1}.
Of particular interest are intrinsic SOIs (ISOIs),  
due to their controllability by tuning
a gate voltage \cite{rashba_book_1,nagaosa_1,winklwer_1,awschalom_1,Miron_1,Erlingsson,tanaka_2}.
Two
commonly studied ISOIs are the Rashba and Dresselhaus types.
The Rashba SOI \cite{rashba_book_1} can be described via spatial
inversion asymmetries, while the Dresselhaus SOI \cite{dresselhaus_term}
is a result of
bulk inversion asymmetries within the crystal structure \cite{winklwer_1,rashba_book_1}.

There  have also been extensive efforts
to manipulate the  spin
currents\cite{prinz_1,wolf_1,zhang_1,kikkawa_1,Demidov}
in SOI systems
via
the spin Hall effect
\cite{Murakami_1,hirsh_1,kato_1,Sinova_1,mishchenko_1,chazalviel,
Nikolic,malsh_severin}, and the quantum spin Hall
effect \cite{Hasan_1,Qi_1}. 
Since
spin currents
are weakly sensitive to nonmagnetic impurities and temperature \cite{zhang_1}, more
opportunities arise in the
development of high speed low-dissipative spintronic devices \cite{kikkawa_1}.
Along these lines,
superconducting heterostructures
have been making strides as potential platforms where
spin-orbit coupling (SOC) plays a key role,
including scenarios involving the spin-Hall effect \cite{malsh_sns,malsh_severin,malsh_3,reynoso_1,ex_so_JJ,Konschelle,yokoyama,tanaka_2,bobkova_1,arahata_1,buzdin_so_ext,golubov_rmp,buzdin_rmp,niu_1,wu_1,bergeret_so}.
When considering superconducting hybrids with SOC, 
interface phenomena at superconducting junctions
becomes particularly important.
For example, the interface of a hybrid superconducting junction
can  behave as a
spin-polarizer
when it is coated by an ultrathin uniform $F$
layer.
The study of interface effects that involve
spin-dependent scattering \cite{spnactv_1,spnactv_6,spnactv_8} 
has spanned 
considerable theoretical \cite{spnactv_12,spnactv_6,spnactv_1,spnactv_21,spnactv_16,spnactv_12,spnactv_14,spnactv_15,spnactv_16,spnactv_20,spnactv_21} 
and
experimental works \cite{Tedrow,DOS_1_ex,DOS_2_ex}.
Advancements in nanofabrication and theoretical techniques
involving 
superconducting hybrids with spin-active
interfaces 
%and intrinsic 
%SOC \cite{bergeret_so,Murakami_1,hirsh_1,kato_1,Sinova_1,mishchenko_1,chazalviel,Nikolic,malsh_sns,malsh_severin,Konschelle},
have thus created new venues for controlling superconducting pair
correlations, spin currents, and majorana fermions 
\cite{spnactv_1,spnactv_6,spnactv_8,spnactv_12,spnactv_14,spnactv_15,spnactv_16,spnactv_20,spnactv_21,nayana}.
%With the recent advancements in nanofabrication technologies and theoretical techniques
%related to 
%superconducting hybrids with spin-active
%interfaces,
%%and intrinsic 
%%SOC \cite{bergeret_so,Murakami_1,hirsh_1,kato_1,Sinova_1,mishchenko_1,chazalviel,Nikolic,malsh_sns,malsh_severin,Konschelle},
%there are incentives to explore
%superconducting systems with %ISOI and 
%interface effects 
%for controlling superconducting pair correlations and spin currents. \cite{spnactv_1,spnactv_6,spnactv_8,spnactv_12,spnactv_14,spnactv_15,spnactv_16,spnactv_20,spnactv_21}

To explore the interplay of these phenomena,
we consider 
charge and spin currents in a
finite sized intrinsic spin-orbit coupled $s$-wave
superconductor/normal-metal/$s$-wave superconductor ($S$/$N$/$S$)
junction with a single spin-active interface.
%We investigate the corresponding
%two-dimensional  $S$/$N$/$S$
%structures 
%using  a quasiclassical method in the diffusive regime.
We utilize a spin-parameterized two-dimensional
Keldysh-Usadel technique \cite{bergeret_so} in the presence of ISOIs.
In order to theoretically
account for spin-polarization and spin-dependent
phase shifts that %an electron 
a quasiparticle experiences upon transmitting across
spin-active interfaces,
spin-boundary conditions are utilized \cite{spnactv_12}.
%The quasiclassical approach was recently
%extended to study diffusive superconducting hybrid structures with
%background spin-dependent fields \cite{bergeret_so}. It was
%demonstrated that a long-range proximity effect can be introduced in
%uniformly magnetized \fs structures due to the momentum-dependence of
%an effective exchange field which relies on the presence of ISOCs in
%the system. This long-range proximity effect is fueled by triplet
%correlations with equal-spin pairings \cite{bergeret1,buzdin_rmp,golubov_rmp},
%and corresponding   $m=\pm 1$ spin projections 
%along the quantization axis.
%Such robust
%proximity-induced superconducting correlations were predicted to
%appear in superconducting hybrids with inhomogeneous ferromagnets or
%uniform magnetic junctions with spin-active
%interfaces \cite{bergeret1,buzdin_rmp,golubov_rmp,halterman1,alid_1,alid_2,alid_3}.
The
spin-parametrization
scheme allows
us to isolate the spin-singlet and spin-triplet correlations, and  pinpoint their
spatial behavior. We find %here 
that the combination of interface spin moment and ISOIs results
in triplet pairings with $m=0,\pm 1$ spin projections 
along the quantization axis \cite{bergeret1,buzdin_rmp,golubov_rmp,halterman1,alid_1,alid_2,alid_3}.
We also find that
maximum singlet-triplet
conversion and spin-current densities
takes place at the corners of the $N$ wire nearest the
spin-active interface, 
where   the  spin accumulation is greatest.
The spin currents possess 
three nonzero spin components, independent of either the actual
type of ISOI present in the \n wire or spin moment orientation of the \ns spin-active interface.
When comparing the spin and charge
currents as functions of the superconducting
phase difference, $\varphi$, we 
%find
show that current phase relations for the charge supercurrents are
typically governed by sinusoidal-like, odd functions in $\varphi$,
although anomalous behavior \cite{buzdin_phi0,yokoyama,Konschelle}
can %occasionally  
arise. The spin currents however, are even functions
of $\varphi$\cite{ma_kh_njp}. These opposing behaviors of the spin and charge
currents present a simple and
experimentally
feasible platform
to
effectively
generate proximity-induced spin-triplet superconducting pairings
and edge spin currents in the absence of any charge supercurrent. It
was demonstrated in Ref. \onlinecite{ma_kh_njp} that the combination of
spontaneously broken time-reversal symmetry and lack of inversion symmetry can result
in spontaneously accumulated spin currents at the edges of finite-size
two-dimensional magnetic \fs
hybrids.
Moreover, we describe
experimentally
accessible signatures in the physically relevant quantities,
and discuss
realistic material parameters and geometrical configurations
that lead to
the
edge spin effects predicted here.
Our
proposed hybrid structure,
based on
its intrinsic properties alone,
can be viewed as a simpler alternative to
differing systems that rely inextricably on externally imposed fields to
generate the desired edge spin
currents \cite{Murakami_1,hirsh_1,kato_1,Sinova_1,mishchenko_1,chazalviel, Nikolic,malsh_sns,malsh_severin}.

The paper is organized as follows: We outline the theoretical
techniques and approximations used to characterize the intrinsic
spin-orbit coupled superconducting \sns hybrid structures with
spin-active interfaces in Sec.~\ref{sec:theor}.
In Sec.~\ref{sec:results},
we discuss all of the types of superconducting pairings present
(spin-singlet and spin-triplet correlations) and illustrate the associated
spatial profiles, which follow directly from the inherent proximity effects.
Next, we present  results for the spin current
densities, with spatial maps, and discuss possible experimental
realizations of the presented edge spin phenomena. In addition,
we discuss the
spin current symmetries relative to the spin moment orientation of
the spin-active interface in the presence of Rashba or Dresselhaus SOC.
We finally summarize our findings in Sec.~\ref{sec:conclusion}.

\section{Theoretical formalism}\label{sec:theor}

A quasiclassical framework has
recently been developed for
superconducting hybrid structures in the presence of generic
spin-dependent fields\cite{bergeret_so}. These generic
fields can be reduced to ISOIs, such as,
Rashba\cite{rashba_term} and Dresselhaus\cite{dresselhaus_term}, in
terms of the quasiparticles' linear momentum [${\bm p}=(p_x,p_y,p_z)$].
If we define a vector
of Pauli matrices ${\bm
\tau}=(\tau_x,\tau_y,\tau_z)$ (see Appendix \ref{app:pauli}),
the corresponding
Hamiltonians describing these
ISOIs can be straightforwardly expressed as,
\begin{eqnarray}\label{rash_dres_Hamiltn}
&&\mathcal{H}_{\text R}=\Omega_{\text R}({\bm p}\times {\bm \tau})\cdot\hat{{\bm z}}\;\eqdef \; \Omega_{\text R}(p_x\tau^y-p_y\tau^x),\nonumber\\
&&\mathcal{H}_{\text D}=\Omega_{\text D}({\bm p}\cdot {\bm
\tau})\;\;\;\;\;\;\;\eqdef \; \Omega_{\text
D}(p_x\tau^x+p_y\tau^y),\nonumber
\end{eqnarray}
where the
momentum is restricted to the $xy$
plane. The ${\text R}$ and ${\text D}$ indices represent
the Rashba and
Dresselhaus SOIs with strength $\Omega_{\text R}$ and $\Omega_{\text
D}$, respectively. A linearized ISOI can be treated as an effective
background field that
obeys the SU(2) gauge symmetries.
\cite{bergeret_so,gorini_1,gorini_3,Konschelle}
Therefore, to incorporate ISOIs into the quasiclassical approach, it
is sufficient
for partial derivatives
to be
interchanged with
their covariants.
\cite{bergeret_so,gorini_1,Konschelle} This
simple prescription
is one of the  advantages of the SU(2)
approach, besides
the convenient definitions of physical quantities such as the spin
currents.\cite{duckheim_1,gorini_3}

A description of quasiparticle transport inside a
superconducting medium is provided by the Dyson
equation. \cite{Eilenberger} The corresponding  equation of motion in the
quasiclassical approximation for clean systems reduces to the
so-called Eilenberger equation\cite{Eilenberger}. The Eilenberger
equation can be further reduced to a simpler set of equations in the
diffusive regime, where the quasiparticles are scattered
into
random
directions.
This permits integration of the Eilenberger equation
over all possible momentum directions,
yielding a simpler
picture for highly impure systems, as
first introduced by Usadel \cite{Usadel}.
\begin{figure}[t!]
\includegraphics[width=8.3cm]{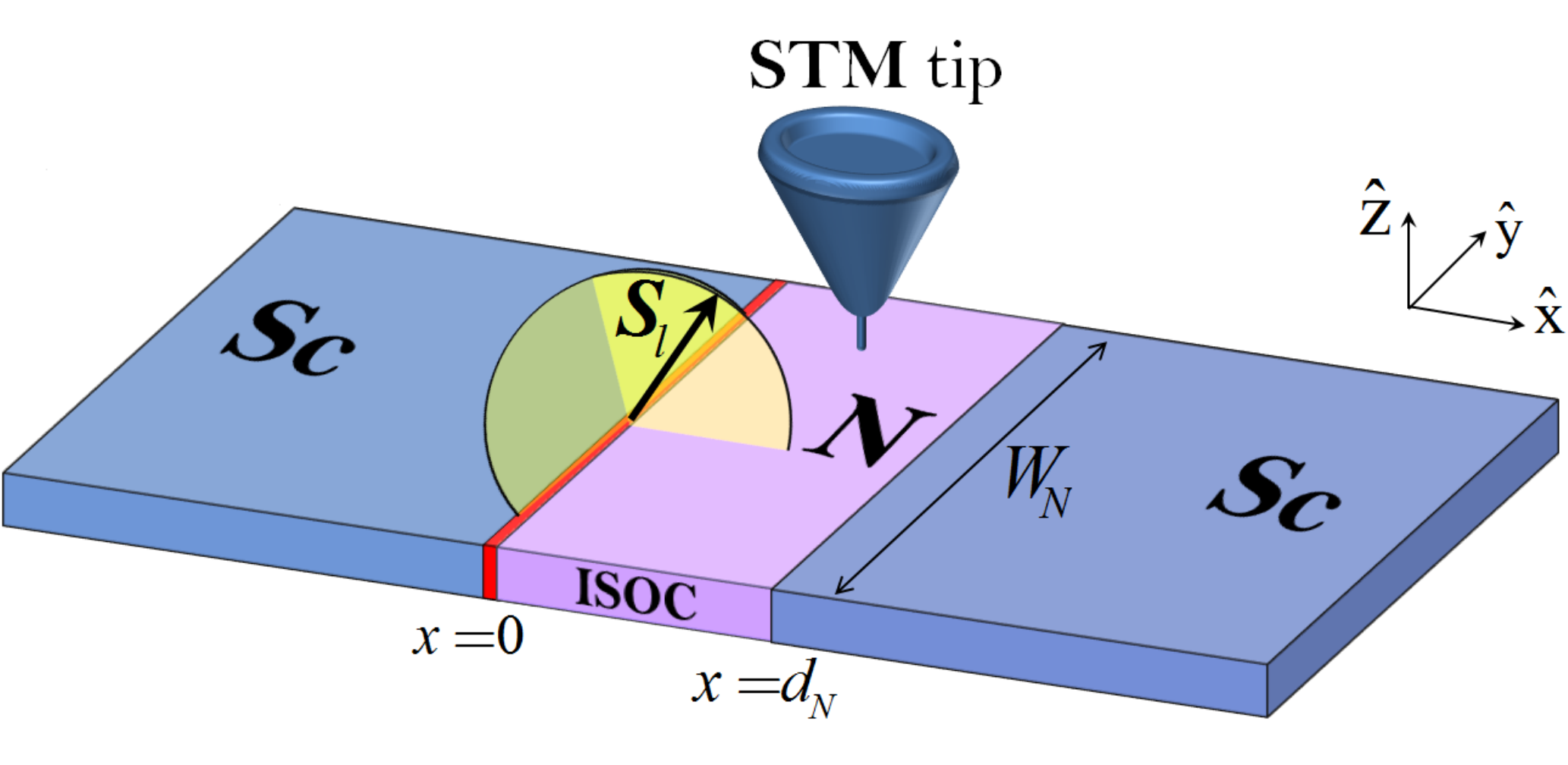}
\caption{\label{fig:model1} (Color online) Schematic of a
two-dimensional \sns Josephson junction with a spin-active interface
at $x=0$ described by ${\bf S}_l=(S_l^x,S_l^y,S_l^z)$. The intrinsic
spin-orbit coupled normal-metal strip ($N$) is of length and width
$d_N$, and $W_N$, respectively. The superconducting electrodes ($Sc$),
however, are SOI-free and connected to the $N$ wire at $x=0$, and $x=d_N$.
The two-dimensional junction resides in the $xy$ plane so that the
\ns interfaces are parallel with the $y$ axis.
The
cone
depicts the tip of a scanning tunneling microscope (STM) that
can sweep the entire
surface of the $N$ layer in the $xy$ plane. }
\end{figure}
The resultant
Usadel equation is the central equation used in this paper,
and can be expressed compactly
as \cite{bergeret1,Usadel,bergeret_so}:
%ma1-> Ref1:cmnt1
\begin{gather}\label{eq:usadel}
 \mathfrak{D}\Big[{\bm \partial},{G}({\bm r},\varepsilon)[{\bm \partial},{G}({\bm
r},\varepsilon)]\Big]+i[ \varepsilon {\rho}_{3}-\Delta,{G}({\bm r},\varepsilon)]=0,
\end{gather}
where ${ \rho}_3$ is a $4\times 4$ Pauli matrix (see Appendix~\ref{app:pauli}), $\mathfrak{D}$ represents the diffusion %ma1-> Ref1:cmnt1-ab %kh diffusive
constant, and $\Delta$ is a $4\times 4$ matrix which represents the superconducting gap \cite{alid_3}. %khf
We
denote the quasiparticle energy by $\varepsilon$, relative
to the Fermi energy, $\varepsilon_F$.
%ma1-> Ref1:cmnt1-ab
In the superconducting leads, the Usadel equation, Eq. (\ref{eq:usadel}), is solved in the presence of $\Delta$ which results in the BCS bulk solution given by Eq. (\ref{eq:bcs}). 
Within the nonsuperconducting region however, $\Delta=0$, and the boundary conditions, Eq. (\ref{eq:bc}), should be simultaneously satisfied. %khf
%ma1 <-
The total Green's function for the system,
$G({\bm r},\varepsilon)$, is comprised of the Advanced ($G^R({\bm r},\varepsilon)$),
Retarded ($G^A({\bm r},\varepsilon)$), and Keldysh ($G^K({\bm
r},\varepsilon)$) propagators:
\begin{eqnarray}
\nonumber && G({\bm r},\varepsilon)=\left[\begin{array}{cc}
           G^R({\bm r},\varepsilon) & G^K({\bm r},\varepsilon) \\
           0 & G^A({\bm r},\varepsilon)
         \end{array}\right];\\\nonumber && G^A({\bm r},\varepsilon)=\left[\begin{array}{cc}
           -\mathcal{G}({\bm r},-\varepsilon) & -\mathcal{F}({\bm r},-\varepsilon) \\
           \mathcal{F}^*({\bm r},\varepsilon) & \mathcal{G}^*({\bm r},\varepsilon)
         \end{array}\right].
\end{eqnarray}
Since we consider the low proximity limit of the diffusive regime,
\cite{bergeret1}, the normal and anomalous components
of the Green's function can be approximated by, $\mathcal{G}({\bm
r},\varepsilon)\simeq 1$ and $\mathcal{F}({\bm r},\varepsilon)\ll
1$, respectively. Thus the advanced component of total
Green's function reduces to:
\begin{eqnarray}\label{gf_F}
{G}^{A}({\bm r},\varepsilon)\approx\left[\begin{array}{cc}
-1 & -\mathcal{F}({\bm r},-\varepsilon)\\
\mathcal{F}^\ast({\bm r},\varepsilon) & 1\\
\end{array}\right].
\end{eqnarray}
Within the low proximity approximation, the advanced component
can be ultimately expressed as:
\begin{eqnarray}\label{Advanced Gree}
&&\nonumber {G}^{A}({\bm r},\varepsilon)=\\&&\left[\begin{array}{cccc}
-1 & 0 & -f_{\upuparrows}({\bm r},-\varepsilon) & -f_{-}({\bm r},-\varepsilon) \\
0 & -1  &-f_{+}({\bm r},-\varepsilon)  & -f_{\downdownarrows}({\bm r},-\varepsilon) \\
 f_{\upuparrows}^{\ast}({\bm r},\varepsilon)  & f_{-}^{\ast}({\bm r},\varepsilon)  & 1  &  0  \\
f_{+}^{\ast}({\bm r},\varepsilon)& f_{\downdownarrows}^{\ast}({\bm r},\varepsilon)  & 0  &  1  \\
\end{array}\right].
\end{eqnarray}
In equilibrium,
the Retarded and Keldysh blocks of the
total Green's function are obtained
via: ${G}^{A}({\bm
r},\varepsilon)=-\big\{\hat{\rho}_3{G}^R({\bm
r},\varepsilon)\hat{\rho}_3\big\}^{\dag}$, and ${G}^{K}({\bm
r},\varepsilon)=\tanh(\varepsilon
k_B\mathcal{T}/2)\big\{{G}^{R}({\bm r},\varepsilon)-{G}^{A}({\bm
r},\varepsilon)\big\}$.  The Boltzmann
constant and system temperature are denoted by
$k_B$ and $\mathcal{T}$, respectively.
Within the low proximity approximation,
the linearized Usadel equation
involves sixteen coupled complex partial differential equations
which become even more complicated in the presence of ISOI terms.
Unfortunately, the resultant system of coupled differential
equations can be simplified and decoupled only in extreme limits
that can be experimentally unrealistic.\cite{golubov_rmp,buzdin_rmp,bergeret1}
When such
simplifications
are made, the equations lead to analytical
expressions for the Green's function
components\cite{alidoust_4termin}. For the complicated
system considered in this paper however, computational methods are the most
efficient,
and sometimes only possible routes to investigate
experimentally accessible transport properties.\cite{bergeret_so}

The
complex partial differential equations
must be
supplemented by
the appropriate boundary conditions to properly
describe the
spin and charge currents
in spin-orbit coupled \sns hybrids with
spin-active interfaces:\cite{cite:zaitsev,spnactv_12}
\begin{eqnarray}\label{eq:bc}
  \nonumber \zeta\big\{{G}({\bm r},\varepsilon){\bm\partial}{G}({\bm r},\varepsilon)\big\}
  \cdot{\bm n}=[{G}_{\text{BCS}}(\theta_{lr},\varepsilon),{G}({\bm
    r},\varepsilon)]\\
    \pm i[{\bm S}_{lr}\cdot{\bm \nu},{G}({\bm r},\varepsilon)],
\end{eqnarray}
where the unit vector,  ${\bm n}$,
is directed normal to an interface, and
 it is assumed for the time being that the left and right interfaces
[Fig.~\ref{fig:model1}] are spin-active.
The leakage intensity of
superconducting correlations from the \s electrodes to the \n wire
is controlled by the ratio between the resistance of
the barrier region $R_B$ and the resistance in the normal
layer $R_N$: $\zeta=R_B/R_N$.\cite{alidoust_1}

We
describe the spin
moments of the left ($l$) and right ($r$) interfaces by two generic
vectors as follows:\cite{spnactv_1,spnactv_6}
\begin{eqnarray}
{\bm S}_{lr}=(S_{lr}^x,S_{lr}^y,S_{lr}^z),
\end{eqnarray}
where the ${\bm S}_{lr}$ can have arbitrary directions and magnitude of the
spin moment at the \ns interfaces.
The solution
to Eq.~(\ref{eq:usadel}) for a bulk even-frequency $s$-wave
superconductor results in
\begin{align}
{G}^{R}_{\text{BCS}}(\theta,\varepsilon)=\left[
                                  \begin{array}{cc}
                                    \sigma^0\cosh\vartheta(\varepsilon) & i\sigma^ye^{i\theta}\sinh\vartheta(\varepsilon) \\
                                    i\sigma^ye^{-i\theta}\sinh\vartheta(\varepsilon) & -\sigma^0\cosh\vartheta(\varepsilon) \\
                                  \end{array}
                                \right],\label{eq:bcs}
\end{align}
where
$\vartheta(\varepsilon)=\text{arctanh}({|\Delta|}/{\varepsilon})$,
and $\theta$ represents the macroscopic phase of the bulk superconductor.
The
phase difference between the left and right \s
electrodes, shown in Fig.~\ref{fig:model1}, is denoted
$\theta_l-\theta_r=\varphi$. We define $s$ and $c$ terms in the
superconducting bulk solution, ${G}_{\text{BCS}}^{R}$, by
piecewise functions:
\begin{eqnarray}
&&\nonumber s(\varepsilon)\equiv e^{i\theta}\sinh\vartheta(\varepsilon)=\\&&\nonumber-\Delta\left\{\frac{\text{sgn}(\varepsilon)}{\sqrt{\varepsilon^2-\Delta^2}}\Theta(\varepsilon^2-\Delta^2)-\frac{i}{\sqrt{\Delta^2-\varepsilon^2}}\Theta(\Delta^2-\varepsilon^2)\right\},\\
&&\nonumber
c(\varepsilon)\equiv\cosh\vartheta(\varepsilon)=\\&&\nonumber\frac{\mid\varepsilon\mid}{\sqrt{\varepsilon^2-\Delta^2}}\Theta(\varepsilon^2-\Delta^2)-\frac{i\varepsilon}{\sqrt{\Delta^2-\varepsilon^2}}\Theta(\Delta^2-\varepsilon^2),
\end{eqnarray}
where $\Theta(x)$ is the Heaviside step function and
$\Delta$ is the superconducting gap at temperature $\mathcal{T}$ for
a conventional $s$-wave superconductor.

We next employ a spin-dependent
field technique that permits the incorporation of ISOIs into the Keldysh-Usadel
approach\cite{bergeret_so}.
As stated earlier, this technique has been widely
used in the literature
\cite{gorini_1,gorini_3,mishchenko_1,malsh_sns,Konschelle}
and was most recently
extended
for superconducting
heterostructures\cite{bergeret_so}. In much the same spirit, we adopt a generic
tensor vector potential ${\bm A}({\bm r})=\big(A_x({\bm r}),A_y({\bm r}),A_z({\bm r})\big)$:\cite{bergeret_so,gorini_1,gorini_3,ma_kh_njp}
\begin{subequations}\label{eq:full_vec_potential}
\begin{align}
A_x({\bm r})=\frac{1}{2}\Big\{\mathcal{A}_x^x({\bm r})\tau^x+\mathcal{A}_x^y({\bm r})\tau^y+\mathcal{A}_x^z({\bm r})\tau^z\Big\},\\
A_y({\bm r})=\frac{1}{2}\Big\{\mathcal{A}_y^x({\bm r})\tau^x+\mathcal{A}_y^y({\bm r})\tau^y+\mathcal{A}_y^z({\bm r})\tau^z\Big\},\\
A_z({\bm r})=\frac{1}{2}\Big\{\mathcal{A}_z^x({\bm
r})\tau^x+\mathcal{A}_z^y({\bm r})\tau^y+\mathcal{A}_z^z({\bm
r})\tau^z\Big\}.
\end{align}
\end{subequations}
We can now define the covariant
derivative by,
\begin{equation}\label{eq:partial_derivativ}
    {\bm\partial}\equiv\vec{\partial} \hat{1}-ie
{\bm A}({\bm r}),
\end{equation}
where $\vec{\partial}\equiv (\partial_x,\partial_y,\partial_z)$.
Hence, the brackets in the Usadel equation Eq. (\ref{eq:usadel}) and
the boundary conditions Eq. ({\ref{eq:bc}}) (as well as the charge
and spin currents discussed below)
are equivalent to:
\begin{equation}\label{eq:brackets}
[{\bm\partial},{G}({\bm r})]=\vec{\partial}{G}({\bm
r})-ie[{\bm A}({\bm r}),{G}({\bm r})].
\end{equation}
It should be noted that the quasiclassical approach employed here
allows for the study %khf
of systems  with spin dependent vector potentials possessing arbitrary spatial %khf
patterns, and spin-active interfaces with arbitrary spin moment
directions.
%ma1-> Ref1:cmnt1-cd
Here we assume that the impurity scattering (encapsulated by the diffusion constant) is spin-independent, and thus the spin-dependent fields introduced in Eqs.~(\ref{eq:full_vec_potential})
describe the spin-orbit coupling for the system. 
Within the quasiclassical approximation, the quasiparticles' momentum is localized around the Fermi level. 
Therefore, the spin moment amplitude for spin-active interfaces 
$|{\bm S}_{l,r}|$, the vector potential,  $|{\bm A}|$, and superconducting gap $|\Delta|$, should all 
be appropriately smaller than the Fermi energy  %khf
$\varepsilon_F$ \cite{bergeret_so}.

A
specific choice of the tensor vector potential\cite{gorini_1}
[Eq.~(\ref{eq:full_vec_potential})] that results in linearized Rashba
($\alpha\neq 0, \beta=0$) \cite{rashba_term} and Dresselhaus
($\beta\neq 0, \alpha=0$) \cite{dresselhaus_term} SOCs is:
\begin{align}\label{eq:Ax_Ay_Az}
    \left\{\begin{array}{l}
      \mathcal{A}_x^x=-\mathcal{A}_y^y=2\beta, \\
      \mathcal{A}_x^y=-\mathcal{A}_y^x=2\alpha, \\
      \hline
      \mathcal{A}_x^z=\mathcal{A}_y^z=0,\\
      \mathcal{A}_z^z=\mathcal{A}_z^x=\mathcal{A}_z^y=0,
    \end{array}\right.
\end{align}
where $\alpha$ and $\beta$ are constants and determine the strength
of Rashba and Dresselhaus SOIs. This choice simplifies the resultant
Usadel equations since
$\vec{\partial}\alpha=\vec{\partial}\beta=0\Rrightarrow\vec{\partial}\cdot\vec{A}({\bm
r})=0$. Hence, by substituting the above set of parameters, Eqs.
(\ref{eq:Ax_Ay_Az}), into Eqs. (\ref{eq:full_vec_potential}) we
arrive at:\cite{bergeret_so}
\begin{subequations}\label{eq:rash_dress}
\begin{align}
A_x=\beta\tau^x-\alpha\tau^y,\\
A_y=\alpha\tau^x-\beta\tau^y.
\end{align}
\end{subequations}
Although these assumptions lead to further simplifications of the Usadel equations,
the end result involves sixteen coupled complex partial differential
equations that we have analytically derived, but omitted here due
to their excessive size.

As mentioned in the introduction,
crystallographic inversion asymmetries\cite{Ganichev} or
lack of structural inversion
symmetries\cite{Miron_1,Garello_1,Duckheim1,Ganichev} in
heterostructures may cause the finite ISOIs considered in this
paper.
For example, engineered strain can induce such inversion
asymmetries\cite{kato_1,cubic_rashba,Nakamura_1,Ganichev} and
consequently, ISOIs. Alternatively, the adjoining of differing
materials may generate interfacial
SOIs\cite{Miron_1,Garello_1,Duckheim1,bergeret_so,Ganichev}.
Nonetheless, currently there is no straightforward method to measure SOIs in a
hybrid structure. One approach might involve first principle
calculations for combined materials. Also, photoemission
spectroscopy\cite{phtoemsn} and spin transfer torque
experiments can provide realistic values for the ISOIs.\cite{bergeret_so,Manchon,Ganichev}.
\begin{figure*}[t!]
\includegraphics[width=18cm,height=7cm]{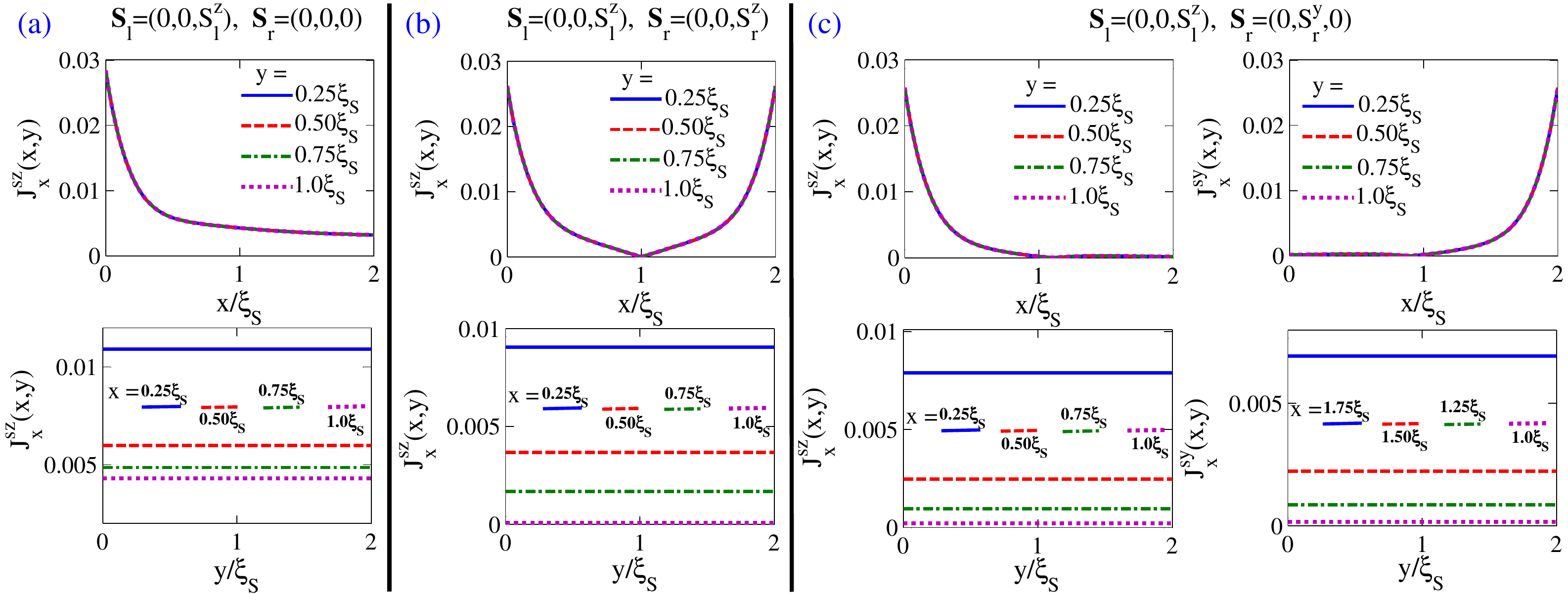}
\caption{\label{fig:spncurrent_so_0} (Color online) Spatial profiles
of the three spin current components, $J^{sx}_x(x,y)$, $J^{sy}_x(x,y)$,
and $J^{sz}_x(x,y)$ in a \sns junction with
spin-active
interfaces and no spin orbit interactions.
Top row panels are against $x$
(positions along the junction length) at differing locations along
$y=0.25\xi_S, 0.5\xi_S, 0.75\xi_S, 1.0\xi_S$ (see Fig.
\ref{fig:model1}). By contrast, the bottom row panels are
functions of $y$, plotted at four positions along the junction
length $x=0.25\xi_S, 0.5\xi_S, 0.75\xi_S, 1.0\xi_S$.
For the $N$ wire we have, $d_N=W_N=2.0\xi_S$,
and the phase difference between the $S$ electrodes is fixed at
$\varphi=\pi/2$. The spin moments of the spin-active interfaces are
denoted by ${\bm S}_{lr}=(S^x_{lr},S^y_{lr},S^z_{lr})$ at the left($l$)
and right($r$) interfaces. (a) The right interface is spin-inactive
while the spin moment of left interface points along the $z$
direction, ${\bm S}_{l}=(0,0,\pm S^z_{l})$. In (b), the two
interfaces are spin-active with parallel spin moments ${\bm
S}_{lr}=(0,0,\pm {S}^z_{lr})$ while in (c), ${\bm S}_{l}=(0,0,\pm
S^z_{l})$ and ${\bm S}_{r}=(0,\pm S^y_{r},0)$.}
\end{figure*}

One of the most striking topics in the
study of transport in junction systems involves spin currents.
Since a decade ago, various features
and behaviors of spin currents in hybrid structures have extensively
been studied.\cite{Nikolic,kato_1,hirsh_1,malsh_sns,
malsh_severin,mishchenko_1,Sinova_1,gorini_3,gorini_1,
Demidov}
The spin and charge currents are key quantities that reveal useful insights
into the system transport characteristics. These physical quantities
are also crucial to the application of nanoscale elements in
superconducting spintronics
devices.\cite{buzdin_rmp,bergeret1}
The vector charge
(${\bm J}^c$) and spin (${\bm J}^{s\gamma}$) current densities can
be calculated by the Keldysh block ($K$) when the system is in
equilibrium:\cite{gorini_1,gorini_3}
\begin{equation}\label{eq:chargecurrentdensity}
{\bm J}^c({\bm r},\varphi) = J_{0}^c
\bigg|\int_{-\infty}^{+\infty}\hspace{-.2cm}
d\varepsilon\text{Tr}\Big\{{\rho}_{3} \big({G}({\bm
r})[{\bm\partial},{G}({\bm r})]\big)^{K}\Big\}\bigg|,
\end{equation}
\begin{equation}\label{eq:spincurrentdensity}
{\bm J}^{s\gamma}({\bm r},\varphi) = J_{0}^s
\bigg|\int_{-\infty}^{+\infty}\hspace{-.2cm}
d\varepsilon\text{Tr}\Big\{{\rho}_{3}
\nu^{\gamma}\big({G}({\bm r})[{\bm\partial},{G}({\bm
r})]\big)^{K}\Big\}\bigg|,
\end{equation}
where $J_{0}^c  =  N_{0} e D/4$, $J_{0}^s=\hbar J_{0}^c/2e$, and
$N_{0}$ is the number of states at the Fermi surface. The vector
current densities provide local directions and amplitudes for the
currents as a function of position. We designate $\gamma=x,y,z$ for
the three components of spin current, e.g., ${\bm J}^{sx}$ represents the
$x$ component of spin current.
The integral of
Eq.~(\ref{eq:chargecurrentdensity}) over the $y$ direction, shown in
Fig. \ref{fig:model1}, provides the total charge supercurrent
flowing across the system.

In order to pinpoint the behavior of the spin-singlet and spin-triplet
pairings inside the spin-orbit coupled $N$ wire, we exploit a
spin-parametrization scheme\cite{buzdin_rmp,bergeret1,alid_4} where the
anomalous component of the Green's function, Eq.~(\ref{gf_F}), can be
parameterized as follows:
\begin{flalign}
&{\mathcal F}({\bm r},\varepsilon)= \mathbb{S}({\bm r},\varepsilon) + {\bm \sigma}\cdot{\bf T}({\bm
r},\varepsilon)= \nonumber\\&\sigma^0\mathbb{S}({\bm r},\varepsilon)+ { \sigma}^{x}{{\text T}}_x({\bm
r},\varepsilon)+ { \sigma}^{y}{{\text T}}_y({\bm
r},\varepsilon)+ { \sigma}^{z}{{\text T}}_z({\bm
r},\varepsilon).
\end{flalign}
In terms of this spin-parametrization, the quantities ${\mathbb S}$ and $T_{x,y}$  
then correspond to the singlet and triplet components with $m=\pm 1$ total spin projection
along the spin quantization axis, while $T_z$ represents the triplet component with $m=0$.\cite{bergeret1,buzdin_rmp,spnactv_6}
Here,
the spin quantization axis is assumed fixed along the $z$ direction
throughout the entire system.
In a uniform ferromagnetic system, it has been demonstrated that the singlet and $m=0$ triplet
component decay rapidly while the
$m=\pm 1$ triplet components, if exist, can propagate over
longer distances compared to the former correlations.
%The long-range triplet correlations have nonzero
%spin-projections along the spin-quantization axis $m=\pm 1$
%whereas the total spin-projection of the short-range component is zero,
%$m=0$.
By noting this aspect of triplet correlations, namely, the degree of their 
penetration (i.e. the distance that the correlations are nonzero)
into a system, one may classify them as `short-range' and
`long-range' correlations. Considering this classification, the $m=0$
triplet component in a uniform ferromagnet is short-ranged
while the $m=\pm 1$ are long-ranged.\cite{bergeret1,buzdin_rmp}
Therefore, the parametrization scheme we utilize allows for
explicit determination of the spatial profiles
for the different superconducting pairings in intrinsically spin-orbit
\sns systems with spin-active interfaces and hence, their short-range and long-range natures.

\section{Results and discussions}\label{sec:results}
For additional insight and comparison purposes, we first consider a \sns junction with
negligible SOIs and
either one or two spin-active interfaces.
We compute the spin currents, and
discuss the singlet and triplet correlations in such systems.
Next, we compute these same quantities after incorporating one of
the Rashba or Dresselhaus ISOIs introduced above.
As remarked earlier, Eq.~(\ref{eq:usadel}) in the presence of
ISOI terms leads to lengthy coupled partial differential equations.
Although
we have analytically derived
the current densities [Eqs.~(\ref{eq:chargecurrentdensity}) and (\ref{eq:spincurrentdensity})]
for numerical  implementation,
they lead to cumbersome
expressions that are not repeated here.
In what follows, we normalize the quasiparticle energy
$\varepsilon$ by the superconducting gap at zero temperature
$\Delta_0=\Delta(\mathcal{T}=0)$, and all lengths by the
superconducting coherence length $\xi_S$ which is defined
as $\sqrt{\hbar \mathfrak{D}/\Delta_0}$. 
%ma1-> Ref2
The barrier resistance at
the \ns interfaces is set to $\zeta=4$. This value of the barrier resistance warrants 
the validity of the low proximity limit i.e. $\mathcal{G}({\bm
r},\varepsilon)\simeq 1$ and $\mathcal{F}({\bm r},\varepsilon)\ll
1$. 
We consider a fixed value for
the spin moment amplitude of the spin-active interfaces, $|{\bm
S}_{lr}|=G_T(\phi_{lr})^{-1}=15$, corresponding to realistic
experimental situations.\cite{Sprungmann}
Here,
$G_T$ is the barrier conductance and 
$\phi_{lr}$ represents the
spin-dependent interfacial phase-shifts at the left and right
interfaces.\cite{spnactv_21,spnactv_12,spnactv_1} 
%ma1-> Ref2
We note that by considering other values for $|{\bm S}_{lr}|$,
within reasonable experimental bounds,
the influence on the results is negligible. %khf
In our actual
computations, we have adopted natural units, so that $k_B=\hbar=1$.
To find stable
solutions to Eq. (\ref{eq:usadel}), and thus to obtain currents
given by Eqs. (\ref{eq:chargecurrentdensity}) and
(\ref{eq:spincurrentdensity}), we have added a small imaginary part,
$\delta\sim 0.01\Delta_0$, to the quasiparticles' energy
$\varepsilon\rightarrow\varepsilon+i\delta$.
Physically, the
additional imaginary part can be considered as a contribution from
inelastic scatterings.\cite{alidoust_1} Due to this imaginary part,
we have taken the modulus of quantities (denoted by the usual
$|...|$). Also, for symmetry reasons we restrict the spatial profiles to the regions
$0<x<d_N/2$ and
$0<y<W_N/2$ in the figures presented throughout the paper.

\subsection{\sns junctions with spin active interfaces in the absence of SOC}

Figure~\ref{fig:spncurrent_so_0} displays the spatial profile of the
spin current components in a \sns junction containing spin-active
interfaces and negligible SOIs.
As seen in Fig.~\ref{fig:model1}, the junction
resides in the $xy$ plane so that the \ns interfaces are located at
$x=0, d_N$ and the vacuum borders in the $y$ direction are found at $y=0, W_N$.
The junction width $W_N$ and length $d_N$ are set to a
representative value of $2.0\xi_S$, without loss of generality.
The top set
of panels in Fig.~\ref{fig:spncurrent_so_0} are functions of the $x$
coordinate at differing $y$, i.e., $0.25\xi_S, 0.5\xi_S, 0.75\xi_S$, and
$1.0\xi_S$, while the bottom row of panels are functions of $y$ at
fixed $x=0.25\xi_S, 0.5\xi_S, 0.75\xi_S$, and $1.0\xi_S$. In
Fig.~\ref{fig:spncurrent_so_0}(a), the spin moment of the left
interface is fixed along $z$, ${\bm S}_l=(0,0,\pm S^z_l)$, while the
right interface is spin-inactive, ${\bm S}_r=0$.
Note that we have denoted vector ${\bm S}$ by three `positive'
scalar entries $S^x, S^y, S^z$, representing projection of the vector in
the $x, y, z$ directions (in the Cartesian coordinate). Here, thus, $\pm$ signs indicate the
orientation of that component which can be parallel (+) or
antiparallel (-) to specific directions $x, y, z$.
It can be seen that
the only nonzero component of spin current is ${\bm J}^{sz}(x,y)$
due to ${\bm S}_l=(0,0,\pm S^z_l)$. This component of spin current
is maximum at the left \ns interface, $x=0$, because of the abrupt
spin-imbalance, and drops to a vanishingly small value near the right
interface, $x=d_N$, which decays to zero over the much longer $d_N$ length scale.
Moreover, the plot demonstrates a uniform distribution of
spin current along $y$. We see that the curves at various
$y$ locations overlap,
consistent with
the bottom panel of Fig.~\ref{fig:spncurrent_so_0}(a), where
the
spin current
is constant for a given
$x$.

%------------------------------------analytics-------------------
As described above and seen in Fig.~\ref{fig:spncurrent_so_0}, the
results for the two-dimensional \sns junction in the absence of ISOIs
reduces it to an effectively one-dimensional  problem. Hence, to gain more insight,
we derive analytical expressions for the charge and spin current
densities in a simple structure, namely a one-dimensional \sns
junction with a single spin-active interface. To this end, we first
derive solutions to the components of the Green's function, Eq.~(\ref{Advanced Gree}), i.e. $f_{\pm}({\bm r},\pm\varepsilon)$,
$[f_{\pm}({\bm r},\pm\varepsilon)]^*$,
$f_{\upuparrows,\downdownarrows}({\bm r},\pm\varepsilon)$, and
$[f_{\upuparrows,\downdownarrows}({\bm r},\pm\varepsilon)]^*$ in a
one-dimensional \sns junction where ${\bm S}_l=(0,0,\pm S^z_l)$ and
${\bm S}_r=0$. If we define a normalized coordinate
$\tilde{x}=x/d_N$, we end up with the following expression for
$f_{-}(x,-\varepsilon)$ (similar expressions arise for the other components):
\begin{eqnarray}
  &&f_{-}(x,-\varepsilon)=\nonumber s^*(\varepsilon)\times\\&&\nonumber
  e^{-i\varphi/2}\frac{\zeta\lambda[\cosh \lambda \tilde{x}+e^{i\varphi}\cosh\lambda(1-\tilde{x})]-iS_l^z\sinh\lambda \tilde{x}}
  {\zeta\lambda[\zeta\lambda\sinh\lambda-iS_l^z\cosh\lambda]}.
\end{eqnarray}
To simplify the solutions, we denote
$2i\varepsilon/\varepsilon_T=\lambda$ (in which $\varepsilon_T$ is the Thouless
energy) and assume $\theta_{lr}=\pm\varphi/2$. By substituting the
obtained solutions into Eq.~(\ref{eq:chargecurrentdensity}) we
arrive at the following expression for the charge current:
\begin{eqnarray}
  &&J_x^c(x,\varphi)=J_0^c\int_{-\infty}^{+\infty}
  d\varepsilon 2i\lambda\tanh(\frac{\varepsilon
  k_BT}{2})\frac{{\cal N}^c}{{\cal D}^c}\sin\varphi,\\&&\nonumber
  {\cal N}^c=\\&&\nonumber{[s^*(-\varepsilon)]}^2\sin\lambda\Big\{{S^z_l}^2-\zeta^2\lambda^2+({S^z_l}^2+\zeta^2\lambda^2)\cosh2\lambda\Big\}\nonumber\\
&&\nonumber+{[s^*(\varepsilon)]}^2\sinh\lambda\Big\{{S^z_l}^2+\zeta^2\lambda^2+({S^z_l}^2-\zeta^2\lambda^2)\cos2\lambda\Big\},\\
&&\nonumber{\cal D}^c=\\&&\nonumber\Big\{
{S^z_l}^2\cos^2\lambda+\zeta^2\lambda^2\sin^2\lambda\Big\} \Big\{
{S^z_l}^2\cosh^2\lambda+\zeta^2\lambda^2\sinh^2\lambda\Big\}.
\end{eqnarray}
It is immediately evident that the charge current has the usual $\sin\varphi$ odd-functionality in the
superconducting phase difference.
The same procedure
is followed to derive analytical expressions for the spin current
components using Eq.~(\ref{eq:spincurrentdensity}).
Since  ${\bm S}_l=(0,0,\pm S^z_l)$, we find
$J^{sx}_x(x,\varphi)=J^{sy}_x(x,\varphi)\equiv 0$ and
$J^{sz}_x(x,\varphi)\neq 0$. Unfortunately, even by means of the
simplifying approximations made to the equations thus far, we arrive at a rather
lengthy expression for the $z$ component of the spin current,
$J^{sz}_x(x,\varphi)$.\cite{ma_kh_njp} Nevertheless, if we restrict our attention
to the edge of the wire,
$x=0$, this spin current component reduces to the following:
\begin{eqnarray}
      J^{sz}_x(\varphi)&=&
    J_0^s\int_{-\infty}^{+\infty} d\varepsilon\frac{4iS^z_l}{\zeta}\tanh(\frac{\varepsilon k_BT}{2})
    \Big\{ \frac{{\cal N}_1^{sz}}{{\cal D}_1^{sz}} +\frac{{\cal N}_2^{sz}}{{\cal
    D}_2^{sz}}\Big\},\\\nonumber
    {\cal N}_1^{sz} &=& [s^*(-\varepsilon)]^2\Big\{2{S^z_l}^2[1+\cos\lambda\cos\varphi]+\zeta^2\lambda^2
    [\cos4\lambda-\\&&\nonumber 6\cos\lambda\cos\varphi-3]+2[{S^z_l}^2+\zeta^2\lambda^2+({S^z_l}^2+3\zeta^2\lambda^2)
    \\&&\nonumber\cos\lambda\cos\varphi]\cos2\lambda\Big\},\\\nonumber
    {\cal N}_2^{sz} &=&[s^*(\varepsilon)]^2\Big\{ 3[{S^z_l}^2+\zeta^2\lambda^2]\cosh\lambda\cos\varphi
    +\\&&\nonumber 4{S^z_l}^2\cosh^2\lambda +({S^z_l}^2-3\zeta^2\lambda^2)\cosh3\lambda\cos\varphi
    \\&&\nonumber -4\zeta^2\lambda^2[2+\cosh2\lambda]\sinh^2\lambda \Big\},
    \\\nonumber
    {\cal D}_1^{sz} &=&\Big\{
    {S^z_l}^2+\zeta^2\lambda^2+[{S^z_l}^2-\zeta^2\lambda^2]\cos2\lambda\Big\}^2,\\\nonumber
    {\cal D}_2^{sz} &=& \Big\{
    {S^z_l}^2-\zeta^2\lambda^2+[{S^z_l}^2+\zeta^2\lambda^2]\cosh2\lambda\Big\}^2.
\end{eqnarray}
The $z$ component of spin current, $J^{sz}_x$, is evidently an even-function of $\varphi$, namely
$\cos\varphi$, although it involves some complicated prefactors.\cite{ma_kh_njp}
%ma1-> Ref2
This finding is consistent with Ref.~\onlinecite{alidoust_1}. Note that %khf
when the other components  of spin current, ${\bm J}^{s\gamma}$, are present,
the even-functionality in $\varphi$ holds, even in the
presence of ISOIs. This effect is discussed further below.
%----------------------------------------------------------------

\begin{figure*}[t!]
\includegraphics[width=18cm,height=6cm]{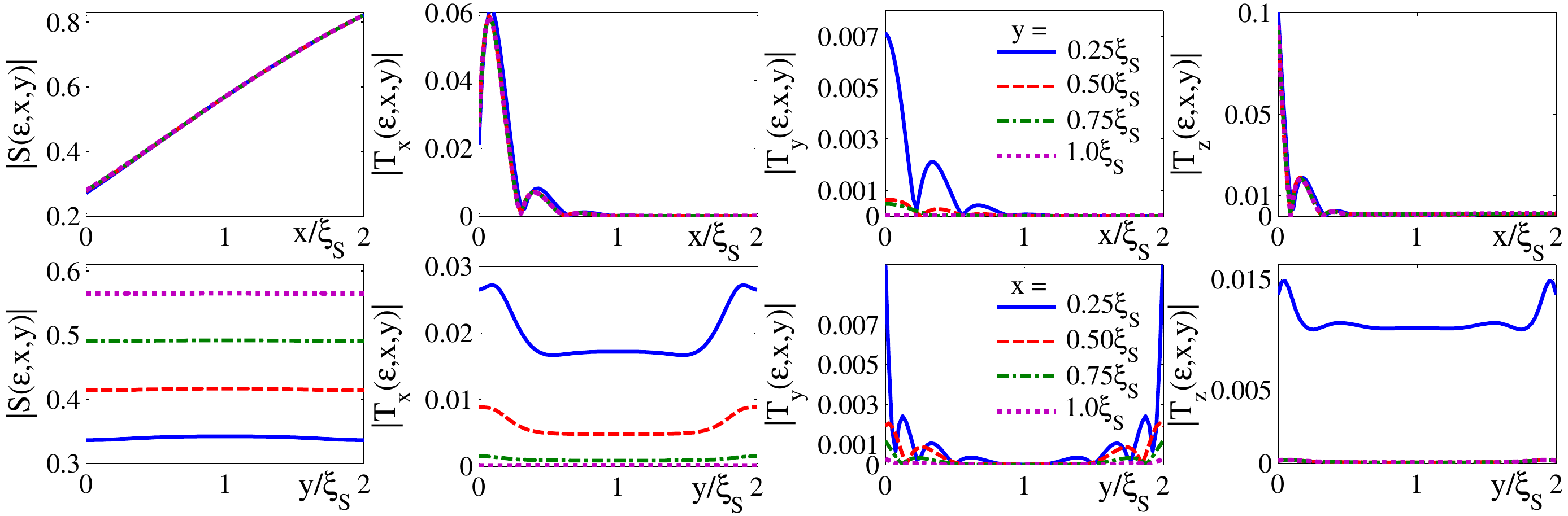}
\caption{\label{fig:triplets} (Color online)
The spatial behaviors of the
singlet $\mathbb{S}(\varepsilon,x,y)$ and triplet ${\bf
T}(\varepsilon,x,y)$ correlations inside a Rashba
\sns system with ${\bf S}_{l}=(0,0,\pm S^z_{l})$ and ${\bf
S}_{r}=0$. The quasiparticles' energy is set at
$\varepsilon=0.1\Delta_0$, $d_N=W_N=2.0\xi_S$, and $\varphi=\pi/2$.
The top panels display the various pair correlations vs $x$ at $y=0.25\xi_S$,
$0.5\xi_S$, $0.75\xi_S$, and $1.0\xi_S$, whereas the bottom panels exhibit the
same quantities as a function of $y$ at $x=0.25\xi_S$, $0.5\xi_S$,
$0.75\xi_S$, and $1.0\xi_S$. }
\end{figure*}

In Fig.~\ref{fig:spncurrent_so_0}(b), the right interface
is now spin-active (at $x=d_N$). The spin moment direction of the
spin-active interface at $x=0$ is intact while ${\bm S}_r=(0,0,\pm
S^z_r)$. As seen in the bottom panel, ${\bm J}^{sz}(x,y)$ is still the only nonzero spin
current component, which is constant in the $y$ direction.
The right spin-active interface causes an increase in  ${\bm J}^{sz}(x,y)$ at
$x=d_N$ due to a spin-imbalance effect.
In Fig.~\ref{fig:spncurrent_so_0}(c), the spin moment
of the interface at $x=d_N$ is oriented along the $y$ direction, i.e., ${\bm
S}_r=(0,\pm S^y_r,0)$. We see that ${\bm J}^{sy}(x,y)$ and ${\bm
J}^{sz}(x,y)$ are both nonzero since ${\bm S}_l$ and ${\bm S}_r$
are orthogonal.
The spin current vanishes at
the middle of the junction ($x=d_N/2$) and apparently the behavior of the
components become interchanged
 at this location. From the bottom row of panels in Fig.~\ref{fig:spncurrent_so_0},
 we see that the spin-active interfaces
with various spin moment orientations would lead to uniformly
distributed spin currents along the junction width in the $y$
direction.
In other words, the spin-active interfaces alone are unable to
induce any spin accumulation at the vacuum borders of the $N$
wire.\cite{Nikolic,kato_1,hirsh_1,
malsh_severin,mishchenko_1,Sinova_1,gorini_3,gorini_1}
The triplet correlations in superconducting hybrids with spin-active
interfaces have extensively been studied.
\cite{spnactv_16,spnactv_12,spnactv_1} In
\sns systems with a single spin-active interface, no equal-spin pairing
can arise (since a single quantization axis exists throughout the whole system),
although opposite-spin triplets, $\text{T}_z(x,y)$, can be induced.
Figure~\ref{fig:spncurrent_so_0}(a), where only $J_x^{sz}$ is nonvanishing,
confirms this phenomenon.

\subsection{Intrinsic spin orbit coupled \sns junctions with a single spin active interface: Singlet and triplet correlations}

Now we incorporate ISOIs in the \sns junction with one spin-active interface
at $x=0$ (see Fig.~\ref{fig:model1}). Figure
\ref{fig:triplets} exhibits first  spatial profiles of the singlet
($\mathbb{S}(\varepsilon,x,y)$) and triplet
($\text{T}_{x,y,z}(\varepsilon,x,y)$) correlations. Here we set
${\bm S}_l=(0,0,\pm S^z_l)$, ${\bm S}_r=0$, and assume that the ISOI
is of the Rashba form, i.e., $\alpha\neq 0$, and $\beta=0$ (we later
discuss the results of a Dresselhaus SOC). We choose
$\alpha=2.0\xi_S^{-1}$\cite{bergeret_so} as a representative value and
emphasize that this specific choice has no influence on the
generality of our findings. In the low proximity limit,
quasiparticles with low energies ($\varepsilon\ll 1$) tend to have
the main
contributions to the pair correlations.
Accordingly, we therefore choose a representative
value for the quasiparticles' energy equal to
$\varepsilon=0.1\Delta_0$.
The other parameters are kept unchanged.
As clearly seen, the combination of an ISOI and spin moment of only
one spin-active interface results in three nonzero components of the
triplet correlations,
which is in contrast to the case with zero ISOI shown in
Fig.~\ref{fig:spncurrent_so_0}. This phenomena directly follows from
the fact that the quasiparticle spin is tied to its momentum in the
presence of an ISOI.\cite{bergeret_so} The singlet
$|\mathbb{S}(\varepsilon,x,y_0)|$ is minimum at $x=0$ and increases
monotonically towards $x=d_N$ at every point along the junction
width $y_0=0.25\xi_S, 0.5\xi_S, 0.75\xi_S$, and $1.0\xi_S$. This
behavior can be understood by noting the combination of interface
spin moment and ISOIs at $x=0$ abruptly converts the singlet
correlations into triplet correlations. This picture is however
reversed at $x=d_N$ where an ingredient to the singlet-triplet
conversion is lacking, i.e., ${\bm S}_r=0$ at $x=d_N$. Examining the
spatial map of the triplet correlations in Fig.~\ref{fig:triplets},
we find that the triplet correlations behave oppositely to the
singlets. The triplet correlations
$\text{T}_{x,y,z}(\varepsilon,x,y_0)$ are maximum near $x=0$, where
the \ns interface is spin-active. Here, as remarked above, the
combination of interface spin moment and ISOIs effectively converts
the singlet superconducting correlations into the triplet ones at
$x=0$. The triplet correlations decline as a function of $x$, and
eventually convert into the singlets at $x=d_N$ (at the
spin-inactive interface). The triplet correlations,
$\text{T}_{x,y}$, have nonzero spin-projections along the
spin-quantization axis ($m=\pm 1$) while $m=0$ for the
$\text{T}_{z}$ component. It is evident that $\text{T}_{z}$ is
drastically suppressed when moving away from the spin-active \ns
interface at $x=0$ (a consequence of the so-called short-range
behavior of $\text{T}_{z}$). The spin-1 quantities,
$\text{T}_{x,y}$, on the other hand, remain nonzero over greater
distances (the so-called long-range behavior).
The actual distances that the triplet correlations can propagate over
them before fully vanishing in a system depends on the system
parameters such as temperature, degree of the interface opacity $\zeta$, strength of the
interface spin-activity and the magnitude of SOCs present in the system. Nonetheless,
a direct comparison of penetration depth between the triplet correlations
with zero total spin $\text{T}_{z}$ and nonzero total spin i.e.
$\text{T}_{x,y}$ clearly reveals that $\text{T}_{z}$ is the
short-ranged triplet component while $\text{T}_{x,y}$ are
long-ranged in the ISO coupled \sns junction with one spin-active
interface. We note that this conclusion 
generally holds, independent of the representative values chosen.
The bottom set of panels in Fig.~\ref{fig:triplets} illustrates the
superconducting correlations $|\mathbb{S}(\varepsilon,x_0,y)|$, and
$|\text{T}_{x,y,z}(\varepsilon,x_0,y)|$ as functions of $y$-position
along the junction width where $x_0=0.25\xi_S, 0.5\xi_S, 0.75\xi_S$,
and $1.0\xi_S$. The spatial distribution of the singlet correlations
along $y$ are unaffected by the coupling of the ISOIs and interface
spin moment at $x=0$. The singlets,
$|\mathbb{S}(\varepsilon,x_0,y)|$, are constant along $y$, implying
a uniform distribution along the junction width. This however
differs considerably from the spatial behavior
of the triplet correlations: The three triplet components
 $|\text{T}_{x,y,z}(\varepsilon,x_0,y)|$ demonstrate appreciable
accumulation at the transverse vacuum boundaries of the $N$ wire (at $y=0$, and
$y=W_N$).
Also, the results reveal that the maximum singlet-triplet
conversions occur at the corners of the $N$ strip near the spin-active
interface (near $x=0$, $y=0$ and $x=0$, $y=W_N$).\cite{ma_kh_njp} Further
investigations have demonstrated that the maximum singlet-triplet
conversion in such systems generally takes place at the corners of
the $N$ wire near any spin-active interfaces.\cite{ma_kh_njp}
We note that this
finding is generic, robust, and independent of either interface spin
moment direction or the actual type of ISOI considered. Similar
spatial profiles appear when $\alpha=0$, and $\beta\neq 0$,
or equivalently when a
Dresselhaus SOI is considered.
Our numerical investigations have
found that the corresponding Dresselhaus SOI results can be
straightforwardly obtained by making the following replacements in Fig.~\ref{fig:triplets}: $|\mathbb{S}^{\text
R}(\varepsilon,x,y)|=|\mathbb{S}^{\text D}(\varepsilon,x,y)|$,
$|{\text T}^{\text R}_{z}(\varepsilon,x,y)|=|{\text T}^{\text
D}_{z}(\varepsilon,x,y)|$, $|{\text T}^{\text
R}_{x}(\varepsilon,x,y)|=|{\text T}^{\text
D}_{y}(\varepsilon,x,y)|$, and $|{\text T}^{\text
R}_{y}(\varepsilon,x,y)|=|{\text T}^{\text
D}_{x}(\varepsilon,x,y)|$. The symmetries can be easily understood
by considering the symmetries of the spin-dependent fields in Eq.~(\ref{eq:rash_dress}),
resulting in Rashba and Dresselhaus
SOIs.\cite{bergeret_so} The spin current components also show
similar symmetries and we shall discuss them in detail at the end of
this section.

 \begin{figure*}[t!]
\includegraphics[width=18cm,height=7cm]{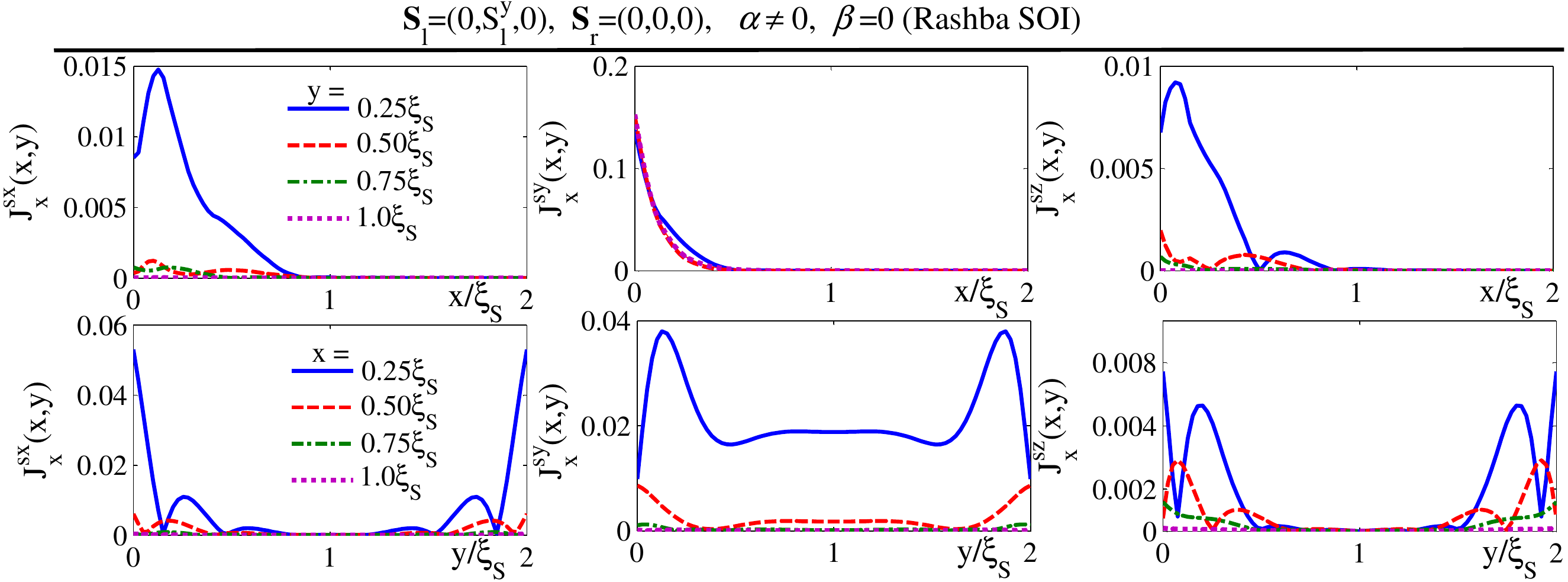}
\caption{\label{fig:snpcrnt_so_rash}
(Color online) The spin current components, $J^{sx}_x(x,y)$,
$J^{sy}_x(x,y)$, $J^{sz}_x(x,y)$ in a Rashba ($\alpha\neq 0,
\beta=0$) \sns junction with one spin-active interface (at $x=0$). The
interface spin moment is oriented along the $y$ direction ${\bf
S}_{l}=(0,\pm S^y_{l},0)$ and ${\bf S}_{r}=0$ (see Fig.
\ref{fig:model1}). The top panels show the spin currents vs $x$ at
$y=0.25\xi_S, 0.5\xi_S, 0.75\xi_S$, and $1.0\xi_S$.  The bottom panels
exhibit the same quantities as a function of $y$ at $x=0.25\xi_S,
0.5\xi_S, 0.75\xi_S$, and $1.0\xi_S$.
The superconducting phase difference
is set at $\varphi=\pi/2$, and $d_N=W_N=2.0\xi_S$.}
\end{figure*}

\subsection{Intrinsic spin orbit coupled \sns junctions with a single spin active interface: Spin currents}

Figure~\ref{fig:snpcrnt_so_rash} exhibits the corresponding spatial
profiles of the spin current components, given by
Eq.~(\ref{eq:spincurrentdensity}) for a Rashba \sns junction with
one spin-active interface. We assume that the left interface at
$x=0$ is spin-active (see Fig.~\ref{fig:model1}), and its spin
moment is oriented along the $y$ direction, namely, ${\bm
S}_l=(0,\pm S^y_l,0)$ (and thus  $ {\bm S}_r=0$). To be specific, we
first present the results of a Rashba ($\alpha\neq 0, \beta=0$) \sns
system in our plots and then later expand our discussion to
differing orientations of ${\bm S}_l$ in the presence of either
Rashba or Dresselhaus SOIs. The top row of panels in
Fig.~\ref{fig:snpcrnt_so_rash} display the three components of spin
current density $J^{s\gamma}_x(x,y_0)$ flowing along $x$ at
$y_0=0.25\xi_S, 0.5\xi_S, 0.75\xi_S$, and $1.0\xi_S$. The spatial
variations of the $x$ component
%[${\bm
%J}^{s\gamma}=(J_{x}^{s\gamma},J_{y}^{s\gamma},J_{z}^{s\gamma})$]
provides the clearest and most useful information
on %to
the spin
current behavior in such systems.
Therefore, we only present the $x$
components in Fig.~\ref{fig:snpcrnt_so_rash}, while later discussing the vector
spin current densities
when presenting symmetries among the spin current components. The
spin current density components vanish within $x>d_N/2$ while they
are largest near the spin-active interface at $x=0$. This behavior
is consistent with the associated triplet correlations investigated
in the top row of panels in Fig.~\ref{fig:triplets}. The spin
currents are zero at $x=d_N$, where the \ns interface is
spin-inactive. This finding is also consistent with previous
theoretical works where the spin currents were found to vanish at
\ns interfaces,\cite{malsh_sns} and thus
there were zero spin currents throughout
the entire ISO coupled \sns hybrids.
Since the \sns system considered here is in an equilibrium state, the
time derivative of the spin density is equal to zero. \cite{gorini_1}
Therefore, because the singlet superconducting electrodes considered
throughout the paper do not support spin currents, the divergence of
the spin current at a \ns interface is zero if the 
spin moment of the spin-active interface is zero (spin-inactive
interfaces).
This fact is clearly seen in Figs.~\ref{fig:spncurrent_so_0}(a) and \ref{fig:snpcrnt_so_rash} where the
right \ns interface is spin-inactive. 
The
bottom panels display the same components except now as a function
of $y$ along the junction width $J^{s\gamma}_x(x_0,y)$ at
$x_0=0.25\xi_S, 0.5\xi_S, 0.75\xi_S$, and $1.0\xi_S$. Most notably, the
plots reveal a nonuniform distribution of spin current
densities along the junction width. From the plots, it is apparent
that the spin current densities peak near the transverse vacuum
boundaries of the $N$ wire at $y=0$, and $y=W_N$. Considering the
top and bottom panels together, we conclude that the spin currents
are maximally accumulated at the edges of the $N$ wire near the
spin-active interface and approximately confined within $x<d_N/2$.
This in turn implies that the corners of the $N$ wire near the
spin-active interfaces possess the maximum of spin current
densities.\cite{ma_kh_njp} The edge accumulation of spin current densities are
reminiscent of those previously found in nonsuperconducting
mesoscale junctions with ISOIs.
\cite{Murakami_1,hirsh_1,kato_1,Sinova_1,mishchenko_1,chazalviel,
Nikolic} We emphasize that the previous works relied
critically on externally applied magnetic and electric
fields\cite{Murakami_1,hirsh_1,kato_1,Sinova_1,mishchenko_1,chazalviel,
Nikolic,malsh_sns,malsh_severin} which can complicate
the theoretical and experimental situations. In contrast, our
findings provide an alternate, simple platform which relies
merely on the intrinsic properties of the system and is devoid of any
externally imposed conditions. Our numerical approach allows us to
determine the precise nature of the spin and charge currents when
varying the superconducting phase difference, $\varphi$. The spatial
maps of charge supercurrent density (not shown) are constant vs
position within the $N$ wire, reflecting the charge conservation
law. We have found that the charge supercurrent is governed by a
sinusoidal-like current phase relation,
while the spin currents, on the contrary, are even-functions\cite{alidoust_1}, i.e.,
${\bm J}^{s\gamma}(\varphi+2\pi)={\bm
J}^{s\gamma}(-\varphi)$. The behavior of charge
supercurrent in the low proximity limit
considered here can differ from the ballistic regime where anomalous
supercurrent-phase relations were found
\cite{Konschelle,yokoyama,reynoso_1}. These findings offer an
appealing experimental platform to examine pure edge spin currents,
not accompanied by charge currents, solely by modulating $\varphi$
without imposing an external electromagnetic field on the system.

A spin-active \ns interface can be ordinarily fabricated by coating
a superconductor with a spin-active layer.
\cite{spnactv_14,spnactv_8,spnactv_15,
spnactv_20,spnactv_1} The signatures of triplet
pairings can be experimentally probed by means of tunneling
experiments and scanning tunneling microscopy/spectroscopy (STM/STS)
techniques\cite{DOS_measur_th}, which rely on zero-energy
peaks in the proximity-induced density of
states\cite{spnactv_14,DOS_1_ex,DOS_2_ex}. Technological
progress allows for measuring high resolution spatially and energy
resolved density of states on a two-dimensional
surface.\cite{DOS_1_ex,DOS_2_ex}
Therefore, the accumulation of triplet
correlations at the boundaries or corners of an $N$ wire, and also
their long-range signatures predicted here, may be realized
in tunneling experiments.
Indeed, one such possibility is  shown
in the schematic of Fig.~\ref{fig:model1},
where the STM tip can traverse the surface and effectively
measure the local
density of states of the entire $N$ layer residing in the $xy$ plane,
producing a spatially-resolved and energy-resolved density of
states\cite{DOS_1_ex,DOS_2_ex}.
Based on our findings described thus far, one can expect
significant modifications to the local density of states as the STM
tip moves toward the edges of the $N$ layer and  probes the
signatures of triplet pairings, which manifest themselves in
zero-energy peaks of the local density of states.

\begin{table}[t]
\begin{center}
\begin{tabular}{|l c|c|c|c|c|c|c|c|c|}
\hline
\multicolumn{1}{|m{2.30cm}}{Magnetic moment ${\bm S}_l$=$(S^x_l,S^y_l,S^z_l)$ ${\bm S}_r$=$(0,0,0)$}&\multicolumn{3}{|c}{${\bm S}_l$=$(\pm S^x_l,0,0)$}&\multicolumn{3}{|c}{${\bm S}_l$=$(0,\pm S^y_l,0)$}&\multicolumn{3}{|c|}{${\bm S}_l$=$(0,0,\pm S^z_l)$}\\
\hline
\multicolumn{1}{|c}{\multirow{2}{*}{Rashba SOI}} & \multicolumn{1}{|c}{\multirow{2}{*}{${\bm J}^{sx}$}} & \multicolumn{1}{|c}{\multirow{2}{*}{${\bm J}^{sy}$}} &  \multicolumn{1}{|c}{\multirow{2}{*}{${\bm J}^{sz}$}} & \multicolumn{1}{|c}{\multirow{2}{*}{${\bm J}^{sx}$}} & \multicolumn{1}{|c}{\multirow{2}{*}{${\bm J}^{sy}$}} &  \multicolumn{1}{|c}{\multirow{2}{*}{${\bm J}^{sz}$}}& \multicolumn{1}{|c}{\multirow{2}{*}{${\bm J}^{sx}$}} & \multicolumn{1}{|c}{\multirow{2}{*}{${\bm J}^{sy}$}} &  \multicolumn{1}{|c|}{\multirow{2}{*}{${\bm J}^{sz}$}} \\[0.4cm]
\hline
\multicolumn{1}{|m{2.30cm}}{Magnetic moment ${\bm S}_l$=$(S^x_l,S^y_l,S^z_l)$ ${\bm S}_r$=$(0,0,0)$}&\multicolumn{3}{|c}{${\bm S}_l$=$(0,\pm S^y_l,0)$}&\multicolumn{3}{|c}{${\bm S}_l$=$(\pm S^x_l,0,0)$}&\multicolumn{3}{|c|}{${\bm S}_l$=$(0,0,\pm S^z_l)$}\\
\hline
\multicolumn{1}{|c}{\multirow{2}{*}{Dresselhaus SOI}} & \multicolumn{1}{|c}{\multirow{2}{*}{${\bm J}^{sy}$}} & \multicolumn{1}{|c}{\multirow{2}{*}{${\bm J}^{sx}$}} &  \multicolumn{1}{|c}{\multirow{2}{*}{${\bm J}^{sz}$}} & \multicolumn{1}{|c}{\multirow{2}{*}{${\bm J}^{sy}$}} & \multicolumn{1}{|c}{\multirow{2}{*}{${\bm J}^{sx}$}} &  \multicolumn{1}{|c}{\multirow{2}{*}{${\bm J}^{sz}$}}& \multicolumn{1}{|c}{\multirow{2}{*}{${\bm J}^{sy}$}} & \multicolumn{1}{|c}{\multirow{2}{*}{${\bm J}^{sx}$}} &  \multicolumn{1}{|c|}{\multirow{2}{*}{${\bm J}^{sz}$}}   \\[0.4cm]
\hline
\end{tabular}
\end{center}
\caption{\label{tab:sym} Symmetries of the spin current components,
${\bm J}^{s\gamma}(x,y)$, in an intrinsic spin-orbit coupled \sns
junction with one spin-active interface depicted in Fig.
\ref{fig:model1}. The spin moments of the left and right interfaces
are denoted by ${\bm S}_{lr}$ and the spin-orbit coupling is set to
be either purely of the  Rashba ($\alpha\neq 0$, $\beta=0$) or Dresselhaus
($\beta\neq 0$, $\alpha=0$) type. To have succinct notation, the
$(x,y)$-functionality of the spin current components is omitted in the
table. The spin current components in similar columns have identical
modulus behaviors.}
\end{table}

Spin accumulation is a distinctive trademark of the spin-Hall
effect.\cite{malsh_sns} Therefore, the  accumulation of
spin current densities at the edges of the sample, as described in this paper, may be directly
measurable through optical experiments such
as Kerr rotation microscopy\cite{kato_1}. In this scenario
spatial maps of the spin polarizations in the entire $N$ wire can be
conveniently imaged.
A multiterminal device can also be alternatively employed to observe
the signatures of spin currents edge accumulation
\cite{Nikolic,mishchenko_1,latrl}. When lateral leads are
attached near the transverse vacuum edges of a two-dimensional \sns
junction (vacuum boundaries at $y=0$, and $y=W_N$ in
Fig.~\ref{fig:model1}), the accumulated spin densities at the
transverse edges of the $N$ wire can result in spin current
injection into the lateral leads, which in turn may induce a voltage
drop. \cite{Nikolic,mishchenko_1,latrl}

Finally, we discuss the symmetries present among the
components of spin current density by varying the spin moment
orientation of a spin-active interface, shown in
Fig.~\ref{fig:model1}, in a Rashba or Dresselhaus \sns hybrid. In
order to systematically obtain and compare results, we first
consider a Rashba SOI ($\alpha\neq 0, \beta=0$) and rotate ${\bm
S}_l$ while ${\bm S}_r=0$. Thereafter, we iterate the same procedure
when the SOI is purely Dresselhaus ($\alpha=0, \beta\neq 0$).
Table~\ref{tab:sym} summarizes the symmetries among the components
of spin current found through extensive numerical investigations.
In the table, vector currents
are presented, i.e., ${\bm J}^{s\gamma}(x,y)=(J_x^{s\gamma},
J_y^{s\gamma}$, and $J_z^{s\gamma})$. The spin current components
with identical spatial maps reside in similar columns. For example,
${\bm J}^{sx}(x,y)$ in the first row and column is identical to
${\bm J}^{sy}(x,y)$ in the second row and first column. In the top
row, labeled ``Rashba SOI", we consider Rashba SOC and rotate the
spin moment of the left interface ${\bm S}_l$. In the bottom row
(labeled ``Dresselhaus SOI"), however, Dresselhaus SOI is considered
and the spin moment rotations, the same as Rashba case, are
iterated. As seen, when the moment of the spin-active interface
points along the $z$ direction ${\bm S}_l=(0,0,\pm S^z_l)$, ${\bm
J^{sz}}$ shows identical behaviors for either Rashba or Dresselhaus
SOIs. However, ${\bm J^{sx}}$ (and ${\bm J^{sy}}$) in the presence
of Rashba SOI is identical to ${\bm J^{sy}}$ (and ${\bm J^{sx}}$) in
the presence of Dresselhaus SOI. This scenario changes when the
moment of the spin-active interface points along the $x$ or $y$
directions. Similar symmetries to the previous case, i.e. ${\bm
S}_l=(0,0,\pm S^z_l)$, are available, provided that we transform
${\bm S}_l=(\pm S^x_l,0,0)$ to ${\bm S}_l=(0,\pm S^y_l,0)$, and vice
a versa, when considering Rashba or Dresselhaus SOIs. For example,
${\bm J^{sz}}$ in the presence of Rashba SOI and ${\bm S}_l=(\pm
S^x_l,0,0)$ is identical to ${\bm J^{sz}}$ in the presence of
Dresselhaus SOI, provided that ${\bm S}_l=(0,\pm S^y_l,0)$. Under
the same conditions, ${\bm J^{sx}}$ (and ${\bm J^{sy}}$) in the
presence of Rashba SOI is identical to ${\bm J^{sy}}$ (and ${\bm
J^{sx}}$) in the presence of Dresselhaus SOI. The contents of
Table~\ref{tab:sym} can be utilized to deduce the spin current
densities in the presence of Rashba (Dresselhaus) SOI solely by
using the results of a Dresselhaus (Rashba) SOI, without any additional
calculations. Similar transformations  can take place based on other
orientations of ${\bm S}_l$. Thus for example, this prescription can
be used to obtain the spatial maps of spin current densities in a
Dresselhaus \sns junction with ${\bm S}_l=(\pm S^x_l,0,0)$ and ${\bm
S}_r=0$ using the data from the plots presented in Fig.~\ref{fig:snpcrnt_so_rash}.

\section{Conclusions}\label{sec:conclusion}

In conclusion, finite-sized two-dimensional intrinsic spin-orbit
coupled \sns junctions with one spin-active interface in the
diffusive regime are theoretically studied using a quasiclassical
approach together with spin-dependent fields obeying SU(2)
gauge symmetries. We
have computed the singlet and triplet
correlations, and the associated spin currents in a  \sns
system where the interface spin moment can take arbitrary orientations.
Using spatial maps of the
singlet and triplet pair correlations within the
two-dimensional $N$ wire, we demonstrate that the combination of one
spin-active interface and an intrinsic spin-orbit interaction (ISOI)
effectively converts singlet pairs into long-range triplet ones.
Interestingly, the spatial profiles illustrate that the
proximity-induced triplet correlations are nonuniformly
distributed and accumulate at the borders of the
$N$ wire nearest the
spin-active interface. By contrast, the spatial amplitude of
the singlet
correlations is uniform within the spin-orbit coupled $N$ wire.
The
results suggest that the maximum singlet-triplet conversion takes
place at the corners of the $N$ wire nearest the spin-active interface.
The spatial profiles of the associated spin current densities also
demonstrate that the three components of spin currents accumulate the most at the
edges of the $N$ wire. Subsequently, the corners of the
$N$ wire near the
spin-active interface host maximum density of spin currents.\cite{ma_kh_njp}
These results are robust and independent of either the interface
spin moment orientation or the actual type of ISOI. (We note that
rich edge phenomena were theoretically found in finite-size two-dimensional intrinsically spin orbit coupled
\sfs junctions in Ref. \onlinecite{ma_kh_njp}).
We also
determine the behavior of spin
and charge currents
by varying
the
macroscopic phase difference between the \s banks, $\varphi$.
The charge supercurrent is
governed by the usual odd-functionality in $\varphi$, while the spin
currents are even-functions of $\varphi$, i.e.
$J^{s}(\varphi+2\pi)=J^{s}(-\varphi)$.\cite{alidoust_1}
Hence
by
properly calibrating $\varphi$, it is possible
to have pure edge spin
currents without driving charge supercurrents.
We then described
experimentally relevant signatures
and potential experiments
aimed at
realizing the edge phenomena predicted here.
Our
work therefore offers a simple structure consisting of a finite-sized
intrinsic spin-orbit coupled \sns junction with one spin-active
interface to generate various edge phenomena, such as
singlet-triplet conversions,  long-range proximity effects, and
 spin currents in the {\it absence} of externally imposed fields.

\acknowledgments

We would like to thank G. Sewell for helpful discussions on the
numerical parts of this work. We also appreciate
N.O. Birge, and F.S.
Bergeret for useful conversations.
K.H. is supported
in part by ONR and by a grant of supercomputer resources
provided by the DOD HPCMP.

\appendix

\section{Pauli Matrices}\label{app:pauli}
In Sec.~\ref{sec:theor} we introduced the Pauli matrices in the spin
space and denoted them by ${\bm\sigma}=\big(\sigma^x, \sigma^y,
\sigma^z\big)$, ${\bm\tau}=\big(\tau^x, \tau^y, \tau^z\big)$, and
${\bm \nu}=\big(\nu^x, \nu^y, \nu^z\big)$.
\begin{align}
&\sigma^x = \begin{pmatrix}
0 & 1\\
1 & 0\\
\end{pmatrix},\;
\sigma^y = \begin{pmatrix}
0 & -i\\
i & 0\\
\end{pmatrix},\;
\sigma^z = \begin{pmatrix}
1& 0\\
0& -1\\
\end{pmatrix},\;
\sigma^0 = \begin{pmatrix}
1 & 0\\
0 & 1\\
\end{pmatrix}.\nonumber
\end{align}
We also introduced the $4\times 4$ matrices
${\bm\rho}=({\rho}_1, {\rho}_2, {\rho}_3)$:
\begin{align}
\nonumber &{\rho}_1 =
\begin{pmatrix}
0 & \sigma^x\\
\sigma^x & 0 \\
\end{pmatrix},\;
{\rho}_2 =  \begin{pmatrix}
0 & -i\sigma^x\\
i\sigma^x & 0 \\
\end{pmatrix},\;
{\rho}_3 = \begin{pmatrix}
\sigma^0 & 0\\
0 & -\sigma^0  \\
\end{pmatrix}.
\end{align}
Following Ref. \onlinecite{alidoust_1}, we define $\tau^\gamma$,
$\nu^\gamma$, and $\rho_0$ as follows;
\begin{align}
 \tau^\gamma = \begin{pmatrix}
\sigma^\gamma & 0\\
0 & \sigma^\gamma\\
\end{pmatrix},\; \nu^\gamma = \begin{pmatrix}
\sigma^\gamma & 0\\
0 & \sigma^{\gamma\ast}\\
\end{pmatrix},\;\rho_0 = \begin{pmatrix}
\sigma^0 & 0\\
0 & \sigma^0 \\
\end{pmatrix},\nonumber
\end{align}
to unify our notation throughout the paper $\gamma$ stands for
$x,y,z$.

\end{document}